\documentclass[prc,twocolumn,showpacs,preprintnumbers,amsmath,amssymb,superscriptaddress]{revtex4}

\usepackage{epsfig}

\def\deq{\delta q}

\begin{document}
\title{Quadrupole collective inertia in nuclear fission: cranking approximation}
\date{\today}

\author{A. Baran}
\affiliation{Department of Physics and
  Astronomy, University of Tennessee, Knoxville, Tennessee 37996, USA}
\affiliation{Physics Division, Oak Ridge National Laboratory, P.O. Box
  2008, Oak Ridge, Tennessee 37831, USA}
\affiliation{Institute of Physics, University of M. Curie-Sk{\l}odowska,
  ul. Radziszewskiego 10, 20-031 Lublin, Poland}

\author{J.A. Sheikh}
\affiliation{Department of Physics and
  Astronomy, University of Tennessee, Knoxville, Tennessee 37996, USA}
\affiliation{Physics Division, Oak Ridge National Laboratory, P.O. Box
  2008, Oak Ridge, Tennessee 37831, USA}

\author{J. Dobaczewski}
\affiliation{Institute of
  Theoretical Physics, University of Warsaw, ul. Ho\.za 69, 00-681
  Warsaw, Poland}
\affiliation{Department of Physics, P.O. Box 35 (YFL),
FI-40014 University of Jyv\"askyl\"a, Finland}

\author{W. Nazarewicz}

\affiliation{Department of Physics and
  Astronomy, University of Tennessee, Knoxville, Tennessee 37996, USA}
\affiliation{Physics Division, Oak Ridge National Laboratory, P.O. Box
  2008, Oak Ridge, Tennessee 37831, USA}
\affiliation{Institute of
  Theoretical Physics, University of Warsaw, ul. Ho\.za 69, 00-681
  Warsaw, Poland}

\begin{abstract}
  Collective mass tensor derived from the cranking approximation to
  the adiabatic time-dependent Hartree-Fock-Bogoliubov (ATDHFB)
  approach is compared with that obtained in the Gaussian Overlap
  Approximation (GOA) to the generator coordinate method. Illustrative
  calculations are carried out for one-dimensional quadrupole fission
  pathways in $^{256}$Fm. It is shown that the collective mass
  exhibits strong variations with the quadrupole collective
  coordinate. These variations are related to the changes in the
  intrinsic shell structure.  The differences between collective
  inertia obtained in cranking and perturbative cranking
  approximations to ATDHFB, and within GOA, are discussed.
\end{abstract}


\maketitle

\section{Introduction}

Microscopic understanding of nuclear collective dynamics is a
long-term goal of low-energy nuclear theory. Large amplitude
collective motion (LACM), as seen in fission and fusion, provides a
particularly important challenge. Those phenomena can be understood in
terms of many-body tunneling involving the mixing of mean fields with
different symmetries.  We have yet to obtain a microscopic
understanding of LACM that is comparable to what we have for ground
states, excited states, and response functions.

For heavy, complex nuclei, the theoretical tool of choice is the
self-consistent nuclear density functional theory (DFT)
\cite{[Rin80],[Ben03]}. The advantage of DFT is that, while treating
the nucleus as a many-body system of fermions, it provides an avenue
for identifying the essential collective degrees of freedom and
provides an excellent starting point for time-dependent
extensions. The time-dependent Hartree-Fock-Bogoliubov (TDHFB) theory
appears, in principle, to provide a proper theoretical framework to
describe the LACM. However, the main drawback of TDHFB, when applied to
fission, is its inability to describe the quantum-mechanical motion
under the collective barrier.

On the other hand, the adiabatic
approximation to TDHFB (ATDHFB) has been successfully applied to the
LACM
\cite{[Bar78],[Kri74],[Bri76],[Dob81],[Gru82],[Hee82],[Gia80a],[Gia80b],[Yul99],[Lib99]}.
The main assumption behind ATDHFB, well fulfilled in the context of
spontaneous fission, is that the collective motion of the system is
slow compared to the single-particle motion of individual nucleons
\cite{[Rin80],[Naz93c]}. According to the path formulation of the
fission problem \cite{[Ska08]}, ATDHFB provides the best framework to
tackle the problem of nuclear dynamics under the barrier. Another
advantage of ATDHFB is that it provides a connection between the
microscopic many-body theory and phenomenological models based on
collective shape variables.

The main theoretical input for an estimate of fission half-lives is
collective inertia (mass tensor) and collective potential. ATDHFB
provides the best framework to calculate mass tensor \cite{[Ska08]}.
However, in most applications, various approximations are adopted. In
the commonly used cranking expression, for instance, the derivatives
with respect to collective coordinates (i.e., collective momenta) are
evaluated using the perturbation theory, and the Thouless-Valatin
self-consistent terms yielding time-odd fields are neglected. The
resulting collective masses are known to be too small \cite{[Dob81],[Yul99]};
hence it is imperative to go beyond the perturbative cranking
treatment.

In the self-consistent investigations of Ref.~\cite{[Yul99]}, based on
the Gogny energy density functional, collective masses were calculated
by explicitly evaluating the collective-coordinate derivatives
appearing in the ATDHFB mass expression.  The resulting collective
mass obtained in such an approach turned out to exhibit appreciable
variations along the collective path, suppressed in the
perturbative cranking treatment.  Furthermore, they noted that ATDHFB
cranking mass could be an order of magnitude greater than the
perturbative cranking mass.  As noted in Ref.~\cite{[Yul99]}, the
enhanced masses obtained in the improved analysis can significantly
impact the calculated fission lifetimes.

The main goal of this work is to investigate the ATDHFB cranking mass
using the nuclear DFT approach with Skyrme energy functionals.  The
paper is organized as follows. Section~\ref{SecATDHFB} summarizes the
basic ATDHFB expressions for collective inertia obtained in
Ref.~\cite{[Dob81]}.  The approximate cranking, perturbative
cranking, and Gaussian Overlap Approximation (GOA)
formulations are given in Sec.~\ref{appATDHFB}.  The
illustrative examples of calculations are contained in
Sec.~\ref{results}, where the results are presented for
$^{256}$Fm. Finally, the main results are given in
Sec.~\ref{summary}.

\section{ATDHFB Theory}\label{SecATDHFB}

This section contains a brief derivation of the collective mass
tensor in the ATDHFB framework. Although some of the expressions are well documented
in the literature~\cite{[Rin80],[Dob81]}, we repeat them here for the sake of
completeness with particular attention paid to various approximations
involved.

\subsection{Summary of HFB}
We begin with the HFB approach.  In what follows, we use the
same notation as in Ref.~\cite{[Dob96fw]}.  The HFB formalism can be
conveniently expressed in terms of the generalized density matrix,
${\cal R}$, defined as
\begin{equation}
{\cal R} =
\Bigg(
\begin{array}{cc}
\rho \quad & \kappa \\
-\kappa^{\ast } \quad & 1-\rho^{\ast }
\end{array}
\Bigg)\,\,\,\,,
\end{equation}
where $\rho$ and $\kappa$ are the particle and  pairing densities
and ${\cal R}^2={\cal R}$.
The energy variation results in the HFB
equation
\begin{equation}
[{\cal W}, {\cal R}] =0\,\,\,, \label{shfb}
\end{equation}
which can be written as a non-linear
eigenvalue problem:
\begin{equation}\label{hfb}
{\cal W}\left( \begin{array}{cc} A & B^\ast \\ B & A^\ast
\end{array}\right)=
\left( \begin{array}{cc} A & B^\ast \\ B & A^\ast
\end{array}\right)
\left( \begin{array}{cc} E & 0 \\ 0 & -E
\end{array}\right)\,\,\,,
\end{equation}
where
\begin{eqnarray}
{\cal W}=&
\left( \begin{array}{cc}
h-\lambda  & \Delta \\
-\Delta^{\ast}  &-h^{\ast} + \lambda
\end{array}\right) \,\,\, , \label{hfbeq}
\end{eqnarray}
$E$ is a diagonal matrix of quasiparticle energies $E_\mu$,
$\lambda$ is the chemical potential,
and matrices $h$ and $\Delta$  are the
particle-hole and pairing mean-field potentials \cite{[Rin80]}, respectively.

For the sake of comparison with the commonly used BCS formalism, it is quite useful to write the HFB equations in the canonical
representation.  The single-particle canonical wave function
$|\mu\rangle$ can be expanded in  the original
single-particle (harmonic oscillator) basis $|n\rangle$ as
\begin{equation}
|\mu\rangle  = \sum_{n}~D_{n \mu} ~ |n\rangle, \label{canmu}
\end{equation}
where the unitary transformation $D$ is obtained by diagonalizing the
density matrix $\rho$. In the canonical basis, the HFB wave function is
given in a BCS-like form:
\begin{eqnarray}
  \breve{A}_{\mu \nu} &=& u_\mu \delta_{\mu \nu}\,\,\,,\,\,\,
  \breve{B}_{\mu \nu} = s_{\bar \mu}^\ast v_\mu \delta_{{\bar \mu} \nu},
  \label{uvmu} \\
  u_\mu &=& u_{\bar \mu} =
  u_{\mu}^\ast\,\,\,,\,\,\,v_\mu = v_{\bar \mu} = v_{\mu}^\ast,
\end{eqnarray}
where the phase $s_{\mu}$, for the time-even quasiparticle vacuum considered here, is defined through the
time-inversion of the single-particle states
\begin{equation}
  \hat T | \mu \rangle =
  s_{\mu} |\bar \mu\rangle \,\,\,,\,\,\, s_{\bar \mu} = -s_{\mu}\,\,\,.
\end{equation}
In Eq.~(\ref{uvmu}) and in the following, the quantities in the canonical basis are denoted by
symbols with breve accents~\cite{[Dob96fw]}.

The HFB energy matrix $\breve{E}$ in the canonical basis is non-diagonal
and is given by
\begin{equation}
  \breve{E}_{\mu \nu} = \xi^+_{\mu \nu} ~(\breve{h}-\lambda)_{\mu \nu} -
  \eta^+_{\mu \nu}~\breve{\Delta}_{\mu \bar \nu}~s_{\bar \nu}^\ast
  \,\,\,, \label{ecan}
\end{equation}
where
\begin{equation}
  \eta^{\pm}_{\mu \nu} = u_\mu v_\nu \pm u_\nu v_\mu \quad {\rm and} \quad
  \xi^{\pm}_{\mu \nu} = u_\mu u_\nu \mp v_\mu v_\nu \,.
\end{equation}
The diagonal matrix elements of the matrix $\breve{E}_{\mu \nu}$ can be
written as \cite{[Rin80],[Dob96fw]}:
\begin{equation}
\breve{E}_{\mu} \equiv \breve{E}_{\mu \mu} =
\sqrt{(\breve{h}_{\mu \mu} -\lambda)^2 + \breve{\Delta}_{\mu \bar \mu}^2}\,\,\,.
 \label{ecan2}
\end{equation}
Even though the above equation resembles the BCS expression for
quasiparticle energy, it involves $\breve{h}_{\mu \mu}$ and
$\breve{\Delta}_{\mu \bar \mu}$, which are respectively obtained by
transforming the HFB particle-hole and the pairing fields to the
canonical basis via the transformation (\ref{canmu}).  It is only
in the BCS approximation that these quantities can be associated with
single-particle energies and the pairing gap.

\subsection{Summary of ATDHFB}

The ATDHFB approach is an approximation to the time-dependent HFB theory,
wherein it is assumed that the collective velocity of the system is
small compared to the average single-particle velocity of the
nucleons.  The generalized HFB density matrix is expanded around the
quasi-stationary HFB solution ${\cal R}_0$ up to quadratic terms in the collective momentum:
\begin{equation}
{ \cal R} = { \cal R}_0 + { \cal R}_1 + { \cal R}_2\,, \label{rexp}
\end{equation}
with ${\cal R}_1$ being time-odd and ${\cal R}_0$ and ${\cal R}_2$ time-even densities.  The
corresponding expansion for the HFB Hamiltonian reads
\begin{equation}
{ \cal W} = { \cal W}_0 + { \cal W}_1 + { \cal W}_2.
\end{equation}
Employing the  density expansion (\ref{rexp}), the HFB energy can be separated
into the collective kinetic and the potential parts.  In terms of the
density expansion (\ref{rexp}), the kinetic energy is given by
\begin{eqnarray}
{\cal K}
&=& \frac{1}{2} {\rm Tr} ( {\cal W}_0 { \cal R}_2 ) +
\frac{1}{4}{\rm Tr} ( {\cal W}_1 { \cal R}_1 )  \nonumber \\
&=& \frac {i} {4} {\rm Tr} \left( {\dot {\cal R}_0} [{ \cal R}_0,
{ \cal R}_1] \right) - \frac {1} {2} \left( [ {\cal R}_2, { \cal R}_0]
[{ \cal W}_0, { \cal R}_0] \right).
\end{eqnarray}
In the usual ATDHFB treatment, the second term involving ${\cal R}_2$
is neglected, and the kinetic energy can be written in the familiar
form:
\begin{equation}
{\cal K} = \frac {1} {2} {\dot q}^2 {\cal M} \,,
\end{equation}
where the collective mass  is  given by
\begin{eqnarray}
{\cal M} &=& \frac {i} {2 {\dot q}^2} {\rm Tr}
  \biggr({\dot {\cal R}_0} [{\cal R}_0, {\cal R}_1]
  \biggr) \label{mass1}\\
&=& \frac {i} {2 {\dot q}} {\rm Tr} \biggr( \frac {\partial {\cal R}_0}
{\partial q} [{\cal R}_0, {\cal R}_1] \biggr). \label{mass3}
\end{eqnarray}
The trace in the above expression can easily be evaluated in the
quasiparticle basis.  To this end, one can utilize the ATDHFB equation
\cite{[Bar78],[Kri74],[Bri76],[Dob81]}
\begin{equation}
i {\dot {\cal R}_0} = [{\cal W}_0, {\cal R}_1 ] +
[ {\cal W}_1, {\cal R}_0]\,. \label{atdhfb1}
\end{equation}
In the quasiparticle basis, the matrices  ${\cal R}_0, {\cal W}_0, {\cal W}_1, {\cal R}_1$, and
${\dot {\cal R}_0}$ are represented by  the matrices ${\cal G}, {\cal E}_0, {\cal E}_1,
{\cal Z}$, and ${\cal F}$, respectively:
\begin{eqnarray}
{\cal R}_0        &=& {\cal A} {\cal G}   {\cal A}^\dagger \,, \\
{\cal W}_0        &=& {\cal A} {\cal E}_0 {\cal A}^\dagger \,, \\
{\cal W}_1        &=& {\cal A} {\cal E}_1 {\cal A}^\dagger \,, \\
{\cal R}_1        &=& {\cal A} {\cal Z}   {\cal A}^\dagger \,, \\
{\dot {\cal R}_0} &=& {\cal A} {\cal F}   {\cal A}^\dagger \,,
\label{r0dot}
\end{eqnarray}
where
\begin{equation}\label{bogo}
{\cal A}=
\left( \begin{array}{cc} A & B^\ast \\ B & A^\ast
\end{array}\right)
\end{equation}
is the matrix of the Bogolyubov transformation, and
\begin{equation}
{\cal G} =
\left(
\begin{array}{cc}
0 & 0 \\
0 & 1
\end{array}
\right)
\quad , {\rm }\quad
{\cal E}_0 =
\left(
\begin{array}{cc}
E & 0 \\
0 & -E
\end{array}
\right) .
\end{equation}
ATDHFB equation (\ref{atdhfb1}) can now be written in the
quasiparticle basis as
\begin{equation}
i {\cal F} = [{\cal E}_0, {\cal Z} ] +
[{\cal E}_1 \,, {\cal G}] \,. \label{atdhfbq1}
\end{equation}
This $2 \times 2$ matrix equation is, in fact, equivalent \cite{[Dob81]}
to the following equation,
\begin{equation}
  i F = E~Z + Z~E + E_1\,\,\,,
\label{atd1can}
\end{equation}
where the antisymmetric matrices $F$, $Z$, and $E_1$ are
related to ${\cal F}$, ${\cal Z}$, and ${\cal E}_1$:
\begin{eqnarray}
{\cal F} =
\left(
\begin{array} {cc}
0 & F   \\
-F^\ast &   0
\end{array}
\right)
\quad &,& \quad
{\cal Z} =
\left(
\begin{array} {cc}
0 & Z   \\
-Z^\ast &   0
\end{array}
\right)
\label{Zmatrix} \\
{}[{\cal E}_1 \,, {\cal G}] &=&
\left(
\begin{array} {cc}
0 & E_1   \\
-E_1^\ast &   0
\end{array}
\right) .
\end{eqnarray}

In the case of several collective coordinates $\{q_i\}$,
the ATDHFB equation (\ref{atdhfb1}) must be solved for each coordinate,
\begin{equation}
i {\dot q}_i\frac {\partial {\cal R}_0}
{\partial q_i} = [{\cal W}_0, {\cal R}_1^i ] +
[ {\cal W}_1^i, {\cal R}_0]\, , \label{atdhfb2}
\end{equation}
and the collective mass tensor becomes:
\begin{equation}
{\cal M}_{ij} =  \frac {i} {2\dot q_j}
{\rm Tr} \biggr( \frac {\partial {\cal R}_0}
{\partial q_i} [{\cal R}_0, {\cal R}_1^j] \biggr). \label{mass4}
\end{equation}
Then, in terms of the corresponding matrices $F^i$ and $Z^j$, the
collective mass tensor is given by
\begin{equation}\label{precis}
   {\cal M}_{ij} = \frac {i} {2{\dot q_i} {\dot q_j}}
   {\rm Tr} \biggr( F^{i\ast} Z^j - F^i Z^{j\ast} \biggr).
\end{equation}
The  expression (\ref{precis}) for the mass tensor contains the matrix $Z^i$, which is
associated with time-odd density matrix ${\cal R}_1^i$ and can, in
principle, be obtained by solving the HFB equations with time-odd
fields. The time-odd fields have been incorporated in mass-tensor
calculations only in a limited number of cases. For instance, in
Ref.~\cite{[Gia80b]}, time-odd fields have been included in the HF study
with a constraint of cylindrical symmetry. The
time-odd fields have also been incorporated in the HFB study in an
approximate iterative scheme with the collective path based on the
Woods-Saxon potential~\cite{[Dob81]}.

\section{Approximations to ATDHFB}\label{appATDHFB}

This section contains the summary of various commonly used
approximations to the exact ATDHFB expression (\ref{precis}).

\subsection{Cranking approximation}

In most of the studies, the time-odd interaction matrix $E_1$
appearing in Eq.~(\ref{atd1can}) is neglected. In the following, this
approximation will be referred to as the cranking approximation
(ATDHFB-C).  In the absence of the term involving $E_1$, the
$Z$-matrix can be easily obtained in the quasiparticle basis
from the equation:
\begin{equation}
-i F^i_{\mu \nu} = (E_\mu + E_\nu) {Z}^i_{\mu \nu}\,\,\,\label{fquasi}
\end{equation}
and the collective cranking mass tensor is given by:
\begin{equation}
  {\cal M}^C_{ij} = \frac {1} {2 {\dot q_i} {\dot q_j}}
  \sum_{\mu \nu} \frac {\left( F^{i\ast}_{\mu \nu} F^{j}_{\mu \nu} +
    F^{i}_{\mu \nu} F^{j\ast}_{\mu \nu} \right) } {E_\mu + E_\nu}.
  \label{mass10}
\end{equation}
It should be noted that Eq.~(\ref{fquasi}) is diagonal in the quasiparticle
basis and not in the canonical basis.
The essential input to the ATDHFB-C mass tensor (\ref{mass10}) is the
matrix $F$. In the following, $F$  is evaluated in both canonical and
quasiparticle basis.

\subsubsection{Canonical basis}
To begin with, Eq.~(\ref{r0dot}) can be written explicitly in terms of
the HFB eigenvectors:
\begin{eqnarray}\label{Rmatrix}
  & &     {\dot {\cal R}_0} =
  {\dot q} \frac {\partial} {\partial q}
  \Bigg(
\begin{array}{cc}
\rho_0 \quad & \kappa_0 \\
  -\kappa_0^{\ast } \quad & 1-\rho_0^{\ast }
\end{array}
\Bigg)
= \\
& & = \left(
\begin{array} {cc}
   A F B^T - B^\ast F^\ast A^\dagger ~~  &
   A F A^T - B^\ast F^\ast B^\dagger \\
   B F B^T - A^\ast F^\ast A^\dagger ~~  &
   B F A^T - A^\ast F^\ast B^\dagger
\end{array}
\right). \nonumber
\end{eqnarray}
Evaluating the matrix elements of (\ref{Rmatrix}) in the canonical
basis, we obtain
\begin{equation}
  \breve{F}^{i}_{\mu \bar \nu} =
  \frac {s_{\bar \nu}} {(u_\mu v_\nu + v_\mu u_\nu)} {\dot q_i}
  \biggr( \frac {\partial \rho_0} {\partial q_i} \biggr)_{\mu \nu}\,.
 \label{fr}
\end{equation}

By differentiating the HFB equation  $[{\cal W}_0, {\cal R}_0] = 0$
with respect to $q_i$,
the derivative of the density matrix in (\ref{fr}) can be expressed in
terms of the derivatives of the particle-hole and the pairing mean-fields. The
resulting 2$\times$2 matrix equation is
\begin{equation}
\bigg[ {\cal A}^\dagger
  {\dot q_i} \frac {\partial {\cal W}_0} {\partial q_i}
  {\cal A}, {\cal G} \bigg]
+ \bigg[ {\cal E}_0, {\cal F} \bigg] = 0.
\end{equation}
By employing the properties of $h$ and $\Delta$ with respect to time
reversal, we obtain
\begin{eqnarray}
  \bigg( \frac {\partial h^\ast} {\partial q_i} \bigg)_{\mu  \nu} &=&
  s^\ast_{\mu} s_{\nu}
  \bigg( \frac {\partial h} {\partial q_i} \bigg)_{\bar \mu  \bar \nu} \,\,\,, \\
  \bigg( \frac {\partial \Delta^\ast} {\partial q_i} \bigg)_{\mu  \nu} &=&
  s^\ast_{\mu} s^\ast_{\nu}
  \bigg( \frac {\partial \Delta} {\partial q_i} \bigg)_{\bar \mu  \bar \nu},
\end{eqnarray}
and by approximating the HFB energy matrix in the canonical basis by
its diagonal matrix elements,
\begin{eqnarray}
  \breve{E}_{\mu\nu} \approx \delta_{\mu\nu}\breve{E}_{\mu},
\label{CRA-BCS}
\end{eqnarray}
one arrives at an approximate ``BCS-equivalent'' expression for the
matrix elements of $F$ in the canonical basis:
\begin{equation}
\breve{F}^{i}_{\mu\nu} \approx \frac{-\dot q_i}{\breve{E}_\mu+\breve{E}_\nu}
\left[ s_\nu\eta^+_{\mu\nu} (\breve{h}^i-\lambda^i)_{\mu\bar\nu}+\xi^+_{\mu\nu}
(\breve{\Delta}^i)_{\mu\nu}\right],
\label{fw}
\end{equation}
where $x^i\equiv \partial x/\partial q_i$ with $x=\breve h$,
$\breve\Delta$, or $\lambda$. In the following, the results obtained by
using this approximation will be called ATDHFB-C$^{\rm c}$.

Using relations (\ref{CRA-BCS}) and (\ref{fw}), the
collective mass tensor (\ref{mass10}) can now be expressed in terms of
the derivatives of the mean-field potentials with respect to the
collective coordinates $q_i$ and BCS-like quasiparticle energies (\ref{ecan2}),
\begin{equation}
  {\cal M}^{C^{\rm c}}_{ij} \approx \frac {1} {2 {\dot q_i} {\dot q_j}}
  \sum_{\mu \nu} \frac {\left( \breve{F}^{i\ast}_{\mu \nu} \breve{F}^{j}_{\mu \nu} +
    \breve{F}^{i}_{\mu \nu} \breve{F}^{j\ast}_{\mu \nu} \right) } {\breve{E}_\mu + \breve{E}_\nu}.
  \label{mass11}
\end{equation}
In the one-dimensional case, the resulting expression agrees with
that of Ref.~\cite{[Yul99]}.

\subsubsection{Quasiparticle basis}

In order to obtain the expression for matrix $F$ in the quasiparticle
basis, we invert Eq.~(\ref{r0dot}) and write the matrix expression for ${\cal
  F}={\cal A}^\dagger {\dot{\cal R}}_0 {\cal A}$
\begin{widetext}
\begin{equation}
{\cal F} =
\left(
\begin{array}{cc}
A^\dagger {\dot \rho}_0 A + A^\dagger {\dot \kappa}_0 B -
B^\dagger {\dot \kappa}_0^\ast A - B^\dagger {\dot \rho}_0^\ast B & \,\,\,
A^\dagger {\dot \rho}_0 B^\ast + A^\dagger {\dot \kappa}_0 A^\ast -
B^\dagger {\dot \kappa}_0^\ast B^\ast - B^\dagger {\dot \rho}_0^\ast A^\ast  \\

B^T {\dot \rho}_0 A + B^T {\dot \kappa}_0 B -
A^T {\dot \kappa}_0^\ast A - A^T {\dot \rho}_0^\ast B & \,\,\,
B^T {\dot \rho}_0 B^\ast + B^T {\dot \kappa}_0 A^\ast -
A^T {\dot \kappa}_0^\ast B^\ast - A^T {\dot \rho}_0^\ast A^\ast
\end{array}
\right)\,\,\,.
\end{equation}
\end{widetext}
Elements (1,1) and (2,2) of ${\cal F}$ vanish because ${\cal R}_0$ is
projective, ${\cal R}_0^2={\cal R}_0$. Equating the above expression with
Eq.~(\ref{Zmatrix}), we obtain
\begin{equation} -F^{\ast} = B^T {\dot \rho}_0 A + B^T {\dot \kappa}_0 B - A^T
  {\dot \kappa}_0^\ast A - A^T {\dot \rho}_0^\ast B \label{fab}\,\,\,.
\end{equation}
In the following, we evaluate the above expression in the simplex
basis, as the mean-field analysis has been performed by imposing this
symmetry. In this basis, the HFB wave function has the following structure
\begin{equation}
B =
\left(
\begin{array}{cc}
B_+ & 0 \\
0     & B_-
\end{array}
\right) \,\,\,
{\rm and }\,\,\,
A =
\left(
\begin{array}{cc}
0 &  A_+\\
A_-     & 0
\end{array}
\right)\,\,.
\end{equation}
The density matrices acquire the following forms in the simplex basis
\begin{eqnarray}
\rho =
\left(
\begin{array}{cc}
B_+^\ast B_+^T & 0 \\
0     & B_-^\ast B_-^T
\end{array}
\right)
&=&
\left(
\begin{array}{cc}
\rho_+ &  0\\
0     & \rho_-
\end{array}
\right) \,\,\,,\\
\kappa =
\left(
\begin{array}{cc}
0 & B_+^\ast A_-^T \\
B_-^\ast A_+^T & 0
\end{array}
\right)
&=&
\left(
\begin{array}{cc}
0 & \kappa_+\\
\kappa_- & 0
\end{array}
\right)\,\,\,.
\end{eqnarray}
The simplex structure of various terms in Eq.~(\ref{fab}) is given by
\begin{eqnarray}
B^T {\dot \rho}_0 A &=&
\left(
\begin{array}{cc}
0 & B^T_+ {\dot \rho}_{0+} A_+ \\
B^T_-{\dot \rho}_{0-} A_- & 0
\end{array}
\right)\,\,\,,
\nonumber \\
A^T {\dot \rho}^\ast_0 B &=&
\left(
\begin{array}{cc}
0 & A^T_- {\dot \rho}_{0-}^{\ast} B_- \\
A^T_+{\dot \rho}_{0^+}^{\ast} B_+ & 0
\end{array}
\right)
\,\,\,,
\nonumber \\
B^T {\dot \kappa}_0 B &=&
\left(
\begin{array}{cc}
0 & B^T_+ {\dot \kappa}_{0+} B_- \\
B^T_-{\dot \kappa}_{0-} B_+ & 0
\end{array}
\right)
\,\,\,,
\nonumber \\
A^T {\dot \kappa}_0^{\ast} A &=&
\left(
\begin{array}{cc}
0 & A^T_- {\dot \kappa}_{0-}^\ast A_+ \\
A^T_+{\dot \kappa}_{0+}^{\ast} A_- & 0
\end{array}
\right)\,\,\,.
\end{eqnarray}
This yields:
\begin{eqnarray}
-F^{\ast} &=&
\left(
\begin{array}{cc}
0 & F_+ \\
F_- & 0
\end{array}
\right),
\end{eqnarray}
where
\begin{eqnarray}
F_+ &=& B^T_+ {\dot \rho}_  {0+} A_+ - A^T_- {\dot \rho}_  {0-}^{\ast} B_- \nonumber \\
    &+& B^T_+ {\dot \kappa}_{0+} B_- - A^T_- {\dot \kappa}_{0-}^{\ast} A_+ ,         \\
F_- &=& B^T_- {\dot \rho}_  {0-} A_- - A^T_+ {\dot \rho}_  {0+}^{\ast} B_+ \nonumber \\
    &+& B^T_- {\dot \kappa}_{0-} B_+ - A^T_+ {\dot \kappa}_{0+}^{\ast} A_-  .
\end{eqnarray}
Since $F$ is antisymmetric, we have obviously $F_+^T=-F_-$, which is
fulfilled explicitly provided $\kappa_+^T=-\kappa_-$.

\subsubsection{Calculation of derivatives}
The collective mass involves either derivatives of the density
matrices or the mean-field potentials. It should be stressed that
these derivatives must be calculated in the original single-particle
basis $|n\rangle$ as the canonical basis (\ref{canmu}) varies with
$\{q_i\}$. In the following, we show how to evaluate the collective
derivatives in the one-dimensional case of single collective
coordinate, the quadrupole deformation $q$.  To this end, we
approximate the derivative of the density operator $\rho$ or $\kappa$
at a deformation point $q=q_0$ by means of the Lagrange three-point
formula for unequally spaced points $q_0-\deq$, $q_0$, and $q_0+\deq'$
\cite{[Gia80b],[Yul99],[Abr70]}:
\begin{eqnarray}\label{lagr}
\biggr( \frac {\partial \rho} {\partial q} \biggr)_{q=q_0}
&\approx & \frac { -\deq'} { \deq (\deq+\deq') } \rho(q_0-\deq) +
    \frac {\deq -\deq'} {\deq\,\deq'} \rho(q_0) \nonumber\\
& & + \frac { \deq} { \deq' (\deq+\deq') } \rho(q_0+\deq').
\end{eqnarray}
The reason for the use of unequally spaced grid in Eq.~(\ref{lagr}) is
purely numerical: the constrained HFB equations cannot be precisely
solved at a requested deformation point $q$.

The corresponding matrix element in the canonical basis can be
expressed through the matrices $D_{n\nu}$ of the canonical
transformation (\ref{canmu}):
\begin{eqnarray}
\bigg( \frac {\partial \rho} {\partial q} \bigg)_{\mu \nu}
\!\!\!& \approx &\!\!\! \frac { -\deq'} { \deq (\deq+\deq') } \sum_{n_1 n_2}
D^\ast_{n_1 \mu} (\rho(q_0-\deq))_{n_1 n_2} D_{n_2 \nu} \nonumber \\
&+&\!\!\! \frac {\deq -\deq'} {\deq\,\deq'}  v^2_\mu \delta_{\mu\nu} \label{lag3} \\
&+&\!\!\! \frac { \deq} { \deq' (\deq+\deq') } \sum_{n_1 n_2}
D^\ast_{n_1 \mu}(\rho(q_0+\deq'))_{n_1 n_2} D_{n_2 \nu}.\nonumber
\end{eqnarray}
It should be noted that the canonical matrix $D_{n\nu}$ in the above
expression corresponds to the deformation point, $q_0$, at which the
mass is evaluated. Furthermore, as mentioned above, the density
matrices at the three deformation points in (\ref{lagr}) need to be
calculated using the single-particle basis $|n\rangle$ with the same basis
deformation.

\subsection{Perturbative cranking approximation}\label{sec:mass}

The perturbative cranking approximation (ATDHFB-C$^{\rm p}$) has been widely used for the
evaluation of the collective mass tensor. In this approximation, apart
from neglecting the time-odd interaction terms in the ATDHFB equation
and off-diagonal matrix elements of the HFB energy matrix
(\ref{CRA-BCS}), the derivatives are not evaluated explicitly but are
obtained using a perturbative approach.  A complete description of the
perturbative cranking model as applied to the nuclear fission process
can be found in Refs.~\cite{[Bel59],[Bes61],[Nil69],[Sob69],[Bra72]}.

The perturbative cranking expression for the mass tensor is obtained by
approximating the mean-field derivatives in Eq.~(\ref{fr}) by
canonical-basis expressions. For instance, the matrix element of $h^i$
can be approximated by
\begin{equation}
\langle\nu|h^i|\mu\rangle\approx
(\breve{h}_{\mu}-\breve{h}_{\nu})\langle\nu|\partial_i\mu\rangle,
\end{equation}
for $\mu\neq\nu$ which, together with
\begin{eqnarray}
\langle\mu|{h}^i|\mu\rangle  \approx
\partial_i \breve{h}_{\mu} & =  & \breve{h}^i_\mu,\\
\partial_i \breve{\Delta}_{\mu} & = & \breve{\Delta}^i_\mu,
\end{eqnarray}
for $\breve{h}_{\mu}\equiv\breve{h}_{\mu\mu}$ and
$\breve{\Delta}_{\mu}\equiv-\breve{\Delta}_{\mu\bar{\mu}}s_{\bar\mu}^*$,
leads to the following expression for the cranking mass tensor
\begin{equation}
{\cal M}^{C^{\rm p}}_{ij} \approx \sum_{\mu \neq \nu}
  \frac{\langle\mu |h^i|\nu\rangle\langle\nu |h^j| \mu\rangle}
  {(\breve{h}_{\mu}-\breve{h}_{\nu})^2(\breve{E}_\mu+\breve{E}_\nu)}
  (\eta^-_{\mu\nu})^2 + \sum_\mu \frac{F_\mu^i F_\mu^j}{2\breve E_\mu}
\label{eq:mass-approx}
\end{equation}
where
\begin{equation}
  F_\mu^i \equiv F_{\mu\bar\mu}^i =
  -\frac{1}{2\breve{E}_\mu^2} [
  \breve{\Delta}_{\mu} (\breve{h}_\mu^i-\lambda^i) -
  (\breve{h}_{\mu}-\lambda)\breve{\Delta}_{\mu}^i].
 \end{equation}
Assuming a weak state dependence of $\breve{\Delta}_\mu$ \cite{[Sie03]},
neglecting the derivatives of $\Delta$ and $\lambda$ \cite{[Bra72]}, and
using  the  identity
\begin{equation}
\frac{\eta^-_{\mu\nu}}{\eta^+_{\mu\nu}}=
\frac{(\breve{h}_{\mu}-\lambda)\breve{\Delta}_{\nu}-(\breve{h}_{\nu}-\lambda)
\breve{\Delta}_{\mu}}
{\breve{E}_\mu\breve{\Delta}_{\nu}+\breve{E}_\nu\breve{\Delta}_{\mu}},
\label{eq:delta-const}
\end{equation}
one arrives at the following perturbative cranking mass tensor:
\begin{equation}
\label{cramass}
{\cal M}^{C^{\rm p}}_{ij} \approx \sum_{\mu\nu}
\frac{\langle\mu |h^i|\nu\rangle\langle\nu |h^j| \mu\rangle}
{(\breve{E}_\mu+\breve{E}_\nu)^3} (\eta^+_{\mu\nu})^2,
\end{equation}
where the sums run over the whole set of canonical states.  This
expression resembles the standard cranking expression for the
collective mass tensor \cite{[Bel59],[Bes61],[Sob69],[Bra72]}
originally derived for a phenomenological mean field $h$.

\subsection{Gaussian overlap approximation}\label{sec:massGOA}

To compare cranking expressions with those obtained within the GOA, it is
convenient to introduce the $\cal S$ matrices \cite{[Sta89a]}:
\begin{equation}\label{sigmas}
  {\cal S}_{ij}^{(K)}=\sum_{\mu,\nu}
  \frac{\langle\mu |h^i|\nu\rangle\langle\nu |h^j| \mu\rangle}
  {(\breve{E}_\mu+\breve{E}_\nu)^{K}} (\eta^+_{\mu\nu})^2.
\end{equation}
It is immediately seen that for the mass tensor of Eq.~(\ref{cramass})
one has ${\cal M}^C = {\cal S}^{(3)}$.  In the case of GOA, also
assuming weak state dependence of pairing and neglecting the
derivatives of $\lambda$ and $\Delta_{\mu\bar\mu}$, one obtains
\cite{[Goz85],[Sta89a]}
\begin{equation}  \label{eq:goa-mass}
  {\cal M}^{\rm GOA}=
  {\cal S}^{(2)}\left[{\cal S}^{(1)}\right]^{-1}{\cal S}^{(2)}.
\end{equation}
Evaluating the matrix elements of $h^i$ entering Eq.~(\ref{sigmas})
perturbatively, one can express ${\cal S}$ explicitly through the
matrix elements of the constraining field operators $\hat Q_i$:
\begin{equation}
  \label{eq:sigma-moment}
  {\cal S}^{(K)}=\frac{1}{4}\left[{ M}^{(1)}\right]^{-1}{ M}^{(K)}
  \left[{ M}^{(1)}\right]^{-1},
\end{equation}
where the energy-weighted  moments ${M}^{(K)}$ are given  by
\begin{equation}
  \label{eq:moment}
  M^{(K)}_{ij} = \sum_{\mu\nu}
  \frac{\langle\mu|\hat Q_i|\nu\rangle
    \langle\nu|\hat Q_j^\dagger|\mu\rangle}
  {(\breve{E}_\mu+\breve{E}_{\nu})^K} (\eta_{\mu\nu}^+)^2.
\end{equation}

\subsection{Treatment of proton and neutron contributions}\label{sec:masspn}

The above expressions for the mass tensor are valid for one kind of
fermions only.  In the case of the cranking approximation, the total mass
tensor is a sum of neutron and proton contributions:
\begin{equation}
  {\cal M}^C_{\rm total} = {\cal M}^C_{\rm n} + {\cal M}^C_{\rm p}.
\end{equation}
In the GOA, however, the total  inverse inertia $({\cal M}^{\rm GOA}_{\rm total})^{-1}$ for a
composite system is given as a sum of proton and neutron inverse covariant
inertia tensors \cite{[Goz85]}:
\begin{equation}
   \label{eq:97}
   ({\cal M}^{\rm GOA}_{\rm total})^{-1}=
    ({\cal M}^{\rm GOA}_n)^{-1}
    + ({\cal M}^{\rm GOA}_p)^{-1}.
\end{equation}


\section{Results}\label{results}

The illustrative calculations were performed for the nucleus  $^{256}$Fm by using
the  SkM$^*$ energy density
functional~\cite{[Bar82]} in the particle-hole (ph) channel. In the particle-particle (pp)
channel we employed the  density-dependent pairing interaction in the mixed
variant of Refs.~\cite{[Dob01c],[Dob02c]}:
\begin{equation}
  V_\tau(\vec r) = V_{\tau 0}\left(1-\rho(\vec r)/2\rho_0\right)\delta(\vec r)\,,
\end{equation}
where $\tau=n,p$ and $\rho_0=0.16\,$fm$^{-1}$.
To test the accuracy of various
approximations, we carried out both  HF+BCS and HFB calculations.
The pairing interaction strengths, which were
adjusted to reproduce the neutron and proton ground-state pairing gaps in
$^{252}$Fm, are (in  $\mbox{MeV fm}^3$):
\begin{equation}
  V_{n0}=-372.0\,,\quad V_{p0}=-438.0\,.
  \label{pair-str}
\end{equation}

The fission pathways were studied in the previous Ref.~\cite{[Sta09]} using the SkM$^*$-HF-BCS approach with the seniority pairing interaction. It has been found that the SkM$^*$ energy density functional favors  the asymmetric fission pathway  in $^{256}$Fm, and our HFB results are consistent with this result.
The one-dimensional collective pathway,
determined by  the axial quadrupole moment $q=Q_{20}$, was obtained by means of
the HFB solver  HFODD  \cite{[Dob09e]}.

\begin{figure}[htb] \includegraphics[width=0.95\columnwidth]{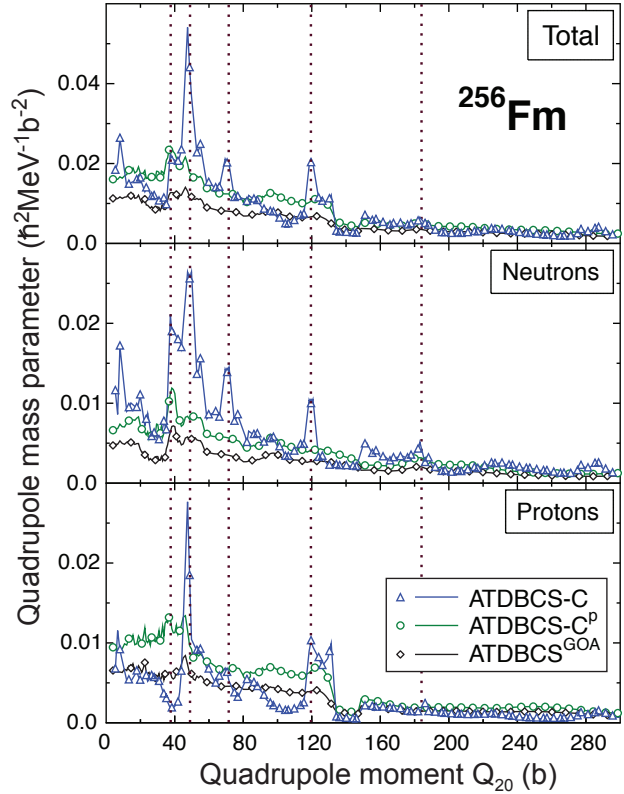}
  \caption{\label{Fig:1} (Color online) The quadrupole  mass parameter (top: total; middle: neutron contribution; bottom: proton contribution)  along the fission pathway of $^{256}$Fm calculated  in SkM$^*$+HF+BCS as a
  function of the mass quadrupole moment.
  The ATDBCS-C results (triangles) are compared with those obtained in the perturbative cranking approximation (ATDBCS-C$^{\rm p}$, circles) and
  Gaussian overlap approximation (ATDBCS$^{\rm GOA}$, diamonds). See text for details.
  }
  \end{figure}
Using the self-consistent solutions along the fission pathway, we calculate the collective quadrupole mass parameter using various
approximations described in Sec.~\ref{appATDHFB}. First, we
discuss results obtained within the HF+BCS formalism (dubbed ATDBCS).
Figure~\ref{Fig:1} compares the results of the non-perturbative cranking approach
(ATDBCS-C) with  the perturbative cranking approximation (ATDBCS-C$^{\rm p}$) and Gaussian overlap approximation (ATDBCS$^{\rm GOA}$). In ATDBCS-C
the  derivatives of the
density matrices and the mean-field potentials  have been obtained using the Lagrange formula, which requires the knowledge of self-consistent
solutions in several neighboring deformation points. We have evaluated the density
matrices for quadrupole deformations ranging from $Q_{20}$= 0 to 320\,b
in steps of 1\,b. The derivatives were obtained by
using the 3-point Lagrange formula (\ref{lag3}), and also
the 5-point Lagrange formula \cite{[Abr70]}. The results for collective mass obtained with 3-point and 5-point expressions differ only in the third decimal
place; hence, in the following, we shall stick to  the 3-point Lagrange
formula. It needs to be stressed  that -- in order
to guarantee consistent labeling of canonical states --
the underlying single-particle basis
should be identical for all three points in Eq.~(\ref{lag3}) involved in the  derivative evaluation.
This has been achieved by performing HF+BCS
calculations using the same basis deformation for all neighboring points.

As seen in Fig.~\ref{Fig:1}, the total
ATDBCS-C mass exhibits a rather irregular behavior characterized by the presence of several sharp maxima.   Some of these peak-like
structures, although considerably suppressed,  also show up in
ATDBCS-C$^{\rm p}$ and ATDBCS$^{\rm GOA}$.

\begin{figure}[htb] \includegraphics[width=0.95\columnwidth]{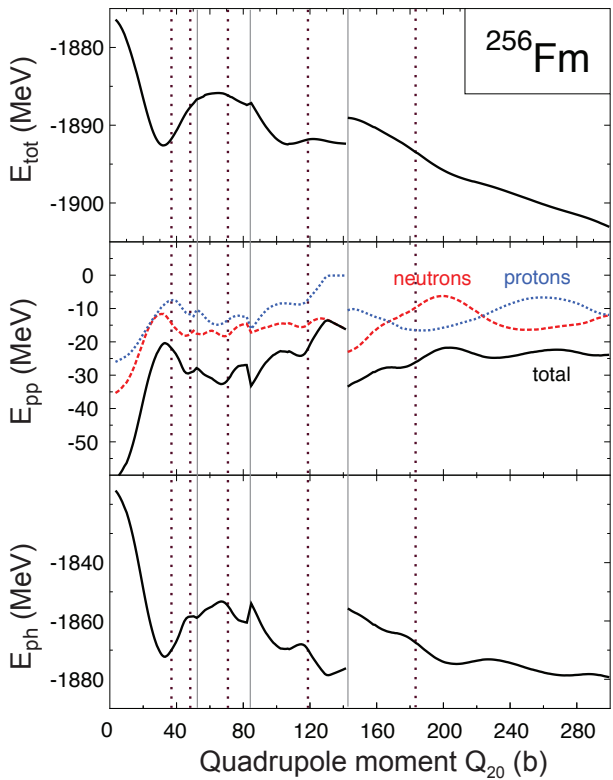}
  \caption{\label{Fig:2} (Color online)
  Total HF+BCS energy $E_{\rm tot}$ (top),
  pairing  energies $E_{\rm pp}$  (total, proton, and neutron; middle), and the HF  energy $E_{\rm ph}$ (bottom) calculated  along the fission pathway for $^{256}$Fm. The borders between different  self-consistent
  configurations are marked by vertical solid lines. The dotted lines mark
  positions of peaks in the collective ATDBCS-C mass parameter of Fig.~\ref{Fig:1}.
  }
 \end{figure}

To unravel the origin of the peak structures in the
collective mass, the total energy of $^{256}$Fm is depicted in
Fig.~\ref{Fig:2}, together with
corresponding pairing-energy
$(E_{\rm pp})$ and HF energy $(E_{\rm ph})$ contributions.
The one-dimensional total energy curve shows several discontinuities due to  intersections of close-lying energy sheets (surfaces) with very different mean fields. The corresponding pathways can in fact be well separated when
studied in more than one  dimension  of the collective manifold \cite{[Sta09]}.
The diabatic jumps between various energy sheets have been disregarded when computing the collective inertia shown in Fig.~\ref{Fig:1}.
Indeed, in such cases the adiabatic theory is unable to provide a
meaningful result for the collective mass.

There are also   configuration changes within each pathway. Because of pairing correlations, these changes are adiabatic in character \cite{[Neg89a],[Naz93c]}.
Still,  they  manifest themselves in the collective inertias through the appearance of peaks \cite{[Gri71a],[Sch80a],[Arv88]}.
Figures \ref{Fig:1} and \ref{Fig:2} nicely illustrate this point: the peaks in the collective ATDBCS-C mass parameter appear in the regions of large local variations in $E_{\rm pp}$ and $E_{\rm ph}$ that are indicative of
changes in the shell structure with elongation. (We note that the local variations in the total energy are much weaker than those in pairing and HF energies, due to the well-known anticorrelation between pairing and HF energies, clearly seen in Fig.~\ref{Fig:2}.)

Two general conclusions can be drawn from the results of Fig.~\ref{Fig:1}. First, the ATDBCS-C$^{\rm p}$ and ATDBCS$^{\rm GOA}$ inertia show fairly similar behavior, with the ATDBCS-C$^{\rm p}$ mass being systematically larger.
Second, the exact treatment of derivative terms in ATDBCS-C gives rise
to less adiabatic behavior in the corresponding collective mass.

\begin{figure}[t] \includegraphics[width=0.95\columnwidth]{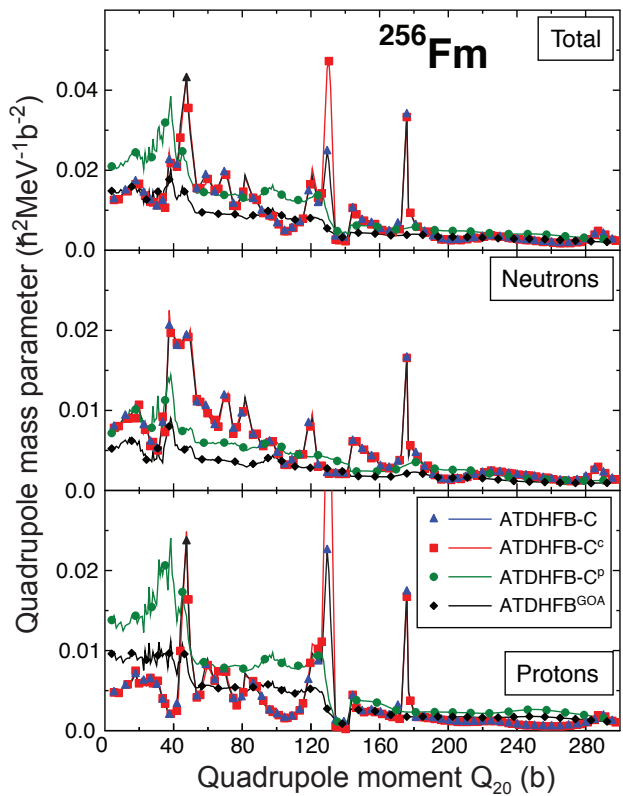}
  \caption{\label{Fig:3} (Color online)  The quadrupole  mass parameter (top: total; middle: neutron contribution; bottom: proton contribution)  along the fission pathway of $^{256}$Fm calculated  in SkM$^*$+HFB as a
  function of the mass quadrupole moment.
  The ATDHFB-C results (triangles) are compared with those obtained in the
  canonical approximation (ATDHFB-C$^{\rm c}$, squares),
  perturbative cranking approximation (ATDHFB-C$^{\rm p}$, dots), and
  Gaussian overlap approximation (ATDHFB$^{\rm GOA}$, diamonds). See text for details.}
  \end{figure}

The results obtained within the  HFB framework are
presented in Fig.~\ref{Fig:3}. The results obtained in the canonical approximation ATDHFB-C$^{\rm c}$  (\ref{mass11}), perturbative treatment of derivatives ATDHFB-C$^{\rm p}$
(\ref{eq:mass-approx}), and ATDHFB$^{\rm GOA}$  (\ref{eq:goa-mass}) were obtained by using the canonical HFB wave functions and employing the diagonal
(``equivalent BCS'')  ansatz. The ATDHFB-C calculations (\ref{mass10}) were carried out  in the full quasiparticle basis.

The most interesting finding is that
the collective mass in ATDHFB-C
is very close to that obtained  in ATDHFB-C$^{\rm c}$. Similar to the HF-BCS
case, the ATDHFB-C$^{\rm p}$ and ATDHFB$^{\rm GOA}$ results follow each other with the ATDHFB-C$^{\rm p}$ mass being systematically larger. Again, the
exact treatment of  derivatives gives rise to less adiabatic behavior of collective mass that manifests itself through the presence of peaks.

In Ref.~\cite{[Yul99]}, the quadrupole collective mass was
evaluated in the canonical basis and exhibited a singular
behavior at certain deformation points.  The primary reason for this
singularity  is due to the pairing
collapse at certain deformations that results in unphysical phase transition and the presence of unavoided level crossings.  In our  work, the
peak structures are  present at nonzero pairing and are related
to the shell structure changes along the fission pathways.

\begin{figure}[htb] \includegraphics[width=0.95\columnwidth]{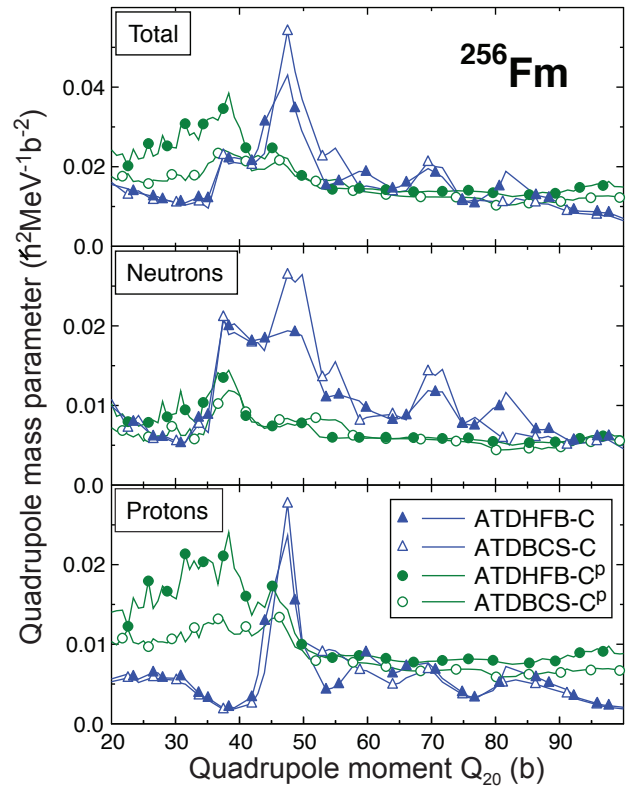}
  \caption{\label{Fig:4} (Color online)
  Similar as in Fig.~\ref{Fig:3} except  for  ATDHFB-C (filled triangles),
  ATDBCS-C (open triangles), ATDHFB-C$^{\rm p}$ (dots), and
 ATDBCS-C$^{\rm p}$ (circles)
   in the narrower region of
  20\,b $\le$$Q_{20}$$\le$100\,b.
  }
  \end{figure}

In order to highlight  the differences between  HFB and
BCS treatments,  Fig.~\ref{Fig:4} shows the quadrupole  masses
obtained in these approaches in the region of the ground-state  minimum and the inner fission barrier of $^{256}$Fm
($Q_{20}\in[20,100]$). This region plays a crucial role in the evaluation of
fission half-lives. It is evident from Fig.~\ref{Fig:4} that
the non-perturbative cranking masses
ATDBCS-C and ATDHFB-C  have very similar behavior.
On the other hand, the masses calculated in the
perturbative approximations, ATDBCS-C$^{\rm p}$ and ATDHFB-C$^{\rm p}$, are quite different for $Q_{20} < 40$. Furthermore, in this region, the perturbative masses appear to be quite large as compared to the cranking values.

The high-frequency fluctuations of collective mass can be traced back to
the imperfect numerical convergence of HFB calculations. In the present work, we assumed
the accuracy of 0.001\,MeV for the total energy. This results in an uncertainty of about 0.002~$\hbar^2$/(MeV b$^2$) in the collective inertia.
If required, the precision of these calculations can  be increased at the
expense of an appreciably higher CPU time.

\section{Summary}\label{summary}

The primary motivation of the present work has been to assess various approximations to the collective mass for fission.
The collective mass plays a crucial  role
in determining the adiabatic collective motion of the nucleus and strongly impacts predicted half-lives. In the majority of previous studies, cranking approximation to collective mass has been employed, in which the time-odd fields are ignored
and the collective momenta  (i.e., derivatives with respect to collective coordinates) needed in the
evaluation of the ATDHFB mass are  calculated using the
perturbation theory.

In our study, we performed the full ATDHFB cranking treatment
of quadrupole inertia.
The numerical evaluation
of the derivatives appearing in ATDHFB mass expression poses a serious computational challenge  as the accurate self-consistent HFB solutions need to be obtained at several neighboring points around every deformation along the fission pathway. By comparing three- and five-point approximations, we conclude   that the three-point Lagrange formula provides a reasonable description of collective derivatives.

The main conclusions of this work can be summarized as follows.
\begin{itemize}
\item
The collective masses obtained in non-perturbative treatment of derivatives
show more variations due to shell structure changes along the fission path as compared to the perturbative approximation and GOA.
\item
The collective mass in full ATDHFB-C
is very close to that obtained  in ATDHFB-C$^{\rm c}$ and ATDBCS-C. This means that the diagonal approximation (\ref{CRA-BCS}) for the  HFB energy matrix
is a very reasonable one.
\item
The ATDHFB-C$^{\rm p}$ and ATDHFB$^{\rm GOA}$ inertias exhibit very similar pattern,  with the ATDHFB-C$^{\rm p}$ mass being systematically larger. A similar conclusion has been reached for the HF+BCS case.
\item
The  main difference between HFB and HF+BCS calculations shows up in
the perturbative treatment: the collective masses calculated in ATDBCS-C$^{\rm p}$ and ATDHFB-C$^{\rm p}$ are sometimes fairly different.
\item
Considering the differences between exact cranking results and
ATDHFB-C$^{\rm p}$ and ATDHFB$^{\rm GOA}$ variants, we conclude that the perturbative treatment of derivatives cannot be justified.
\end{itemize}

The present work deals with the cranking approximation to ATDHFB in which only time-even mean
fields have been kept when  evaluating the  collective inertia. The discussion of the full ATDHFB treatment, including the time-odd response that is expected to play a significant role in the description of collective dynamics \cite{[Mat10]}, will be the subject of a forthcoming study.

\acknowledgments

This work was supported in part by the National Nuclear Security
Administration by the National Nuclear Security Administration under the Stewardship Science Academic Alliances program through DOE Grant
DE-FG52-09NA29461; by the U.S. Department of Energy under Contract
Nos. DE-FG02-96ER40963 (University of Tennessee), and
DE-FC02-09ER41583 (UNEDF SciDAC Collaboration);
by the NEUP grant
DE-AC07-05ID14517 (sub award
00091100); by the Polish Ministry of Science under Contracts
Nos. N~N202~328234 and N202~231137; and by the Academy of Finland
and University of Jyv\"askyl\"a within the FIDIPRO programme.
Computational resources were provided by the National Center for
Computational Sciences at Oak Ridge National Laboratory and the
National Energy Research Scientific Computing


\end{document}